# Imaging the building blocks of ant pheromone communication

**Authors**: Poissonnier L.-A.*, Amichot E., Brevier J., Liard T., Bon P.*
**Affiliation:** XLIM, CNRS UMR 7252, University of Limoges, Limoges, France
*: corresponding authors

**Abstract**

*Organisms have evolved diverse strategies to communicate. Beyond acoustic and visual signals, many species rely on chemical communication, including pheromones: volatile compounds involved in reproduction, defence, and coordination. Ant trail systems, a paradigm of pheromone use, have inspired decades of research in collective behaviour, self-organization and optimization algorithms. Yet, despite their central role, individual pheromone deposits have never been directly visualized or quantified, limiting our understanding of how deposition events scale into collective trail formation. Here, we use interferometric imaging to reveal and quantify single pheromone deposits in the ant* Lasius niger*. This label-free, non-invasive approach enables precise measurement of single deposit volume (≈1 pL, with >20-fold variability), morphology, and temporal dynamics. We uncover a two-phase volume loss process: most of secreted volume evaporates within seconds, forming a dried droplet, which then partially sublimates (≈10%) over one hour. These findings challenge the assumption of homogeneous deposition and bring previously inaccessible properties of ant pheromone trails within experimental reach. Our method establishes a quantitative framework for studying the spatiotemporal dynamics of insect chemical communication, linking individual signalling to collective patterns.*

## Main

Pheromones play central roles across taxa, but are particularly important in insects, where they regulate mating behaviours, spatial orientation, aggregation, response to danger, and caste determination (*1*, *2*). Among insect communication systems, the odour trail system of ants is seen by some authors as the most elaborate of all known forms of chemical communication (*3*). By depositing pheromone droplets from their abdomen, ants create olfactory paths that guide colony members to resources such as food. While the existence of these trails was first documented in 1779 (*4*), their chemical basis remained unidentified for nearly two centuries. The first active components were isolated from leaf-cutting ants, requiring 3.7 kg of workers for characterization (*5*). To date, trail-following pheromones have been identified in over 60 ant species.

Yet fundamental questions persist regarding the spatiotemporal dynamics of pheromone deposition and composition. These gaps limit our understanding of how individual deposits scale into collective trail formation and self-organized networks. As a result, although ant trails have become a key model for self-organization and have extensively inspired the use of so-called Ant Colony Optimisation algorithms (ACO) (*6*, *7*), progress remains limited by the lack of quantitative data needed to parameterize and test theoretical models.

A major obstacle is the lack of direct quantification of individual pheromone deposits, whose volumes and evaporation dynamics remain unknown. Current methods infer trail strength indirectly from ant traffic or deposition behaviour, assuming uniform and continuous pheromone release. However, factors such as food quality (*8*–*12*) and nest distance are known to influence deposition patterns (*13*). One early study using an indirect visualization (soot displacement) suggest heterogeneity in deposits in the ant *Lasius interjectus* (*9*). The assumption of constant deposit volume lacks empirical support, and variability—both within and between individuals—could profoundly affect trail emergence and stability. Without quantification of single deposits, the absolute concentration of natural trails remains unknown, as do critical parameters such as diffusion rates, evaporation dynamics, and detection thresholds.

Synthetic trails, created using known attractive compounds in a solvent, provide controlled conditions but do not reflect the chemical complexity, spatial distribution or even the state of matter of natural deposits. Illustrating this complexity, discrepancies between glandular composition and natural trail content have been observed in the Argentine ant *Linepithema humile* (*14*–*16*). Existing methods lack the resolution to associate chemical composition with individual deposition events, and even the nature of the solvent (likely constituting most of the deposited volume) remains unresolved.

These gaps persist because pheromone deposits have not been directly observed. Recent advances in imaging now offer the possibility of overcoming this limitation. Here, we report the visualization and quantification of trail pheromone deposits in the ant *Lasius niger* using quantitative phase imaging, an interferometric label-free approach that enables live, high-resolution analysis of deposit dynamics.

## Live imaging of a pheromone deposition event

To image a live pheromone deposit event, we imaged a fixed portion of a suspended glass bridge on top of which ants could walk freely from their colony. To trigger pheromone deposition, a sugar reward was placed at the end of the bridge (Fig 1a). Using quantitative phase imaging (Fig 1b), we captured two transparent deposits expulsed from the abdomen during stereotypical pheromone-laying behaviour (Fig 1c-f and Suppl. Movie 1).

Interferometric quantitative phase imaging (*17*) gives access to the spatiotemporal variation of the local phase-shift introduced by the sample on the incident light beam (Fig 1b). This phase-shift can be converted to thickness under the hypothesis of a known refractive index (see methods). The first deposit (Fig 1c,d,g) appeared liquid when ejected from the abdomen and quickly lost about 2/3 of its original volume (0.4 pL to 0.15 pL in about 120 ms). The second deposit (Fig 1e,f,h), produced 2.5 minutes after the first one, was more than fourfold larger in volume and exhibited the same evaporation behaviour, losing about 3/4 of its initial volume (1.7pL to 0.4pL) with an exponential decay of $\tau_{short} = 11$ s. Evaporation led to a roundish static solid structure of approximatively 80 µm diameter and an median thickness of 42 nm for the first deposit, and a more elongated solid structure of 100x50 µm and an median thickness

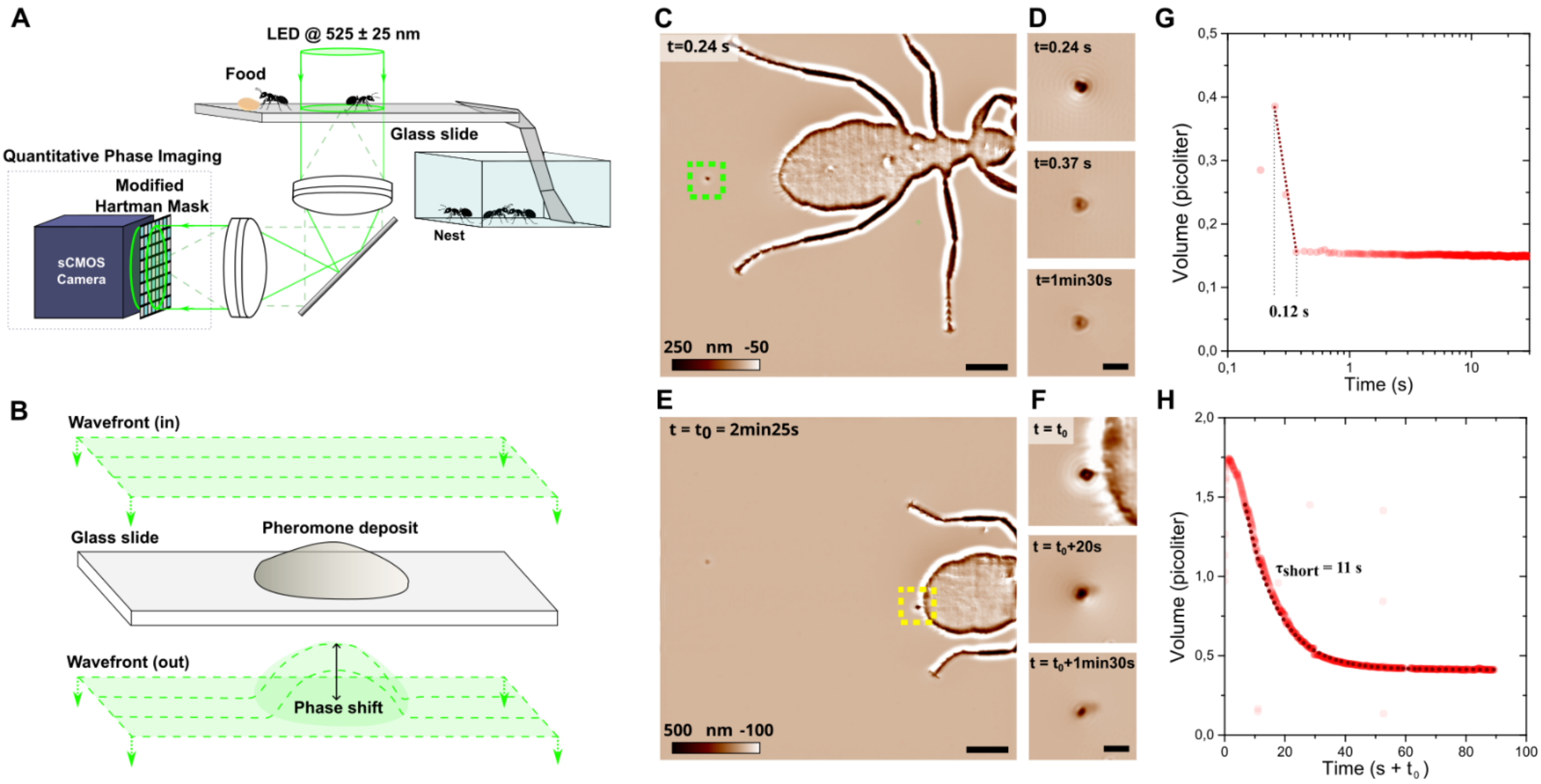

of 78 nm for the second deposit.

**Figure 1: Interferometric quantification of live pheromone deposits**. (a) Schematic representation of the setup to perform live imaging of ant pheromone deposit using quantitative phase optical imaging in transmission. (b) Illustration of the phase-shift introduced by the pheromone deposit. (c-f) quantitative phase images (converted into local thickness) of ants at different time points. Pheromone deposits are visible in the respectively green and yellow squared area. The colour lookup table (LUT) codes for the local thickness. (d) zoom on the green square of panel c at different time scale with the same LUT as c. (g) Total pheromone volume variation over time (in log scale) of the first deposit (c,d) showing a fast evaporation with exponential fit. (h) same as g for the second pheromone deposit in the yellow square (e,f) with linear time scale and exponential fit. Scale bars, 500 µm (c,e), 100 µm (d,f).

## Temporal dynamic of a pheromone deposit over an hour

We then adapted our protocol to monitor individual deposits over longer timescales and at higher spatial resolution. Behavioural studies have established that a longer time scale evaporation should also be present. In *L. niger*, trail pheromones deposited on glass have for instance been reported to elicit a following behaviour for up to 45 min (*18*). Ants were tested individually, the cleanliness of the glass substrate was recorded under the microscope equipped with quantitative phase imaging (see methods) before each trial, to avoid any false positive due contamination from the environment. Because the imaging field of view at high resolution was small (340x340 µm²), the probability of an ant depositing pheromone precisely within the imaged area was low, and deposition events were not recorded live. Instead, ant behaviour was used to guide the localization of putative deposits for subsequent imaging. When laying pheromones, *L. niger* workers decrease their speed and bend their abdomen so it can touch the ground. When we observed this behaviour, we immediately put the glass bridge under the microscope to image the corresponding area until we found the deposit, (ca 30 s after deposition). The deposit (after the fast volume loss observed in Fig 1) had an ovoid shape of 80x40um (Fig 2a,b) with very inhomogeneous thickness ranging from 20 to 500 nm, as highlighted in the line outs presented in Fig 2c along the long (light blue dotted line) and short axis (dark blue dashed line). These values are comparable with Fig 1. The volume of the dried deposit was 0.3 pL. We observed a small (10%) volume lost from the deposit over more than an hour ($\tau_{long}$=63 min), indicating that only part of the droplet was lost, leaving a solid residue. This unexpected finding was confirmed by re-imaging the residue 24h later (Suppl. Info. 1). This suggests that the volatile fraction of the trail detected by the ants is likely very minute compared to what is deposited on the walking substrate. The decay time was of the order of magnitude as the reported duration of pheromone activity known in *L. niger.* Moreover, matter loss was spatially heterogeneous across the droplet ($e_{lost} = 12.2^{+5.4}_{-7.3}$ nm, corresponding to the median and 25%/75% percentiles), revealing denser areas (Fig 2d and vertical histogram Fig 2g and Suppl. Movie 2). In contrast, the decay times were comparatively more homogeneous across the deposit ($\tau_{long} = 67.5^{+4.2}_{-15.5}$ min, Fig 2e,f and horizontal histogram Fig 2g). Considering in addition the very low thickness (≈0.5 µm)/diameter (≈50 µm) ratio, it suggests that the majority of the lost matter corresponds to the same chemical compound across the deposit.

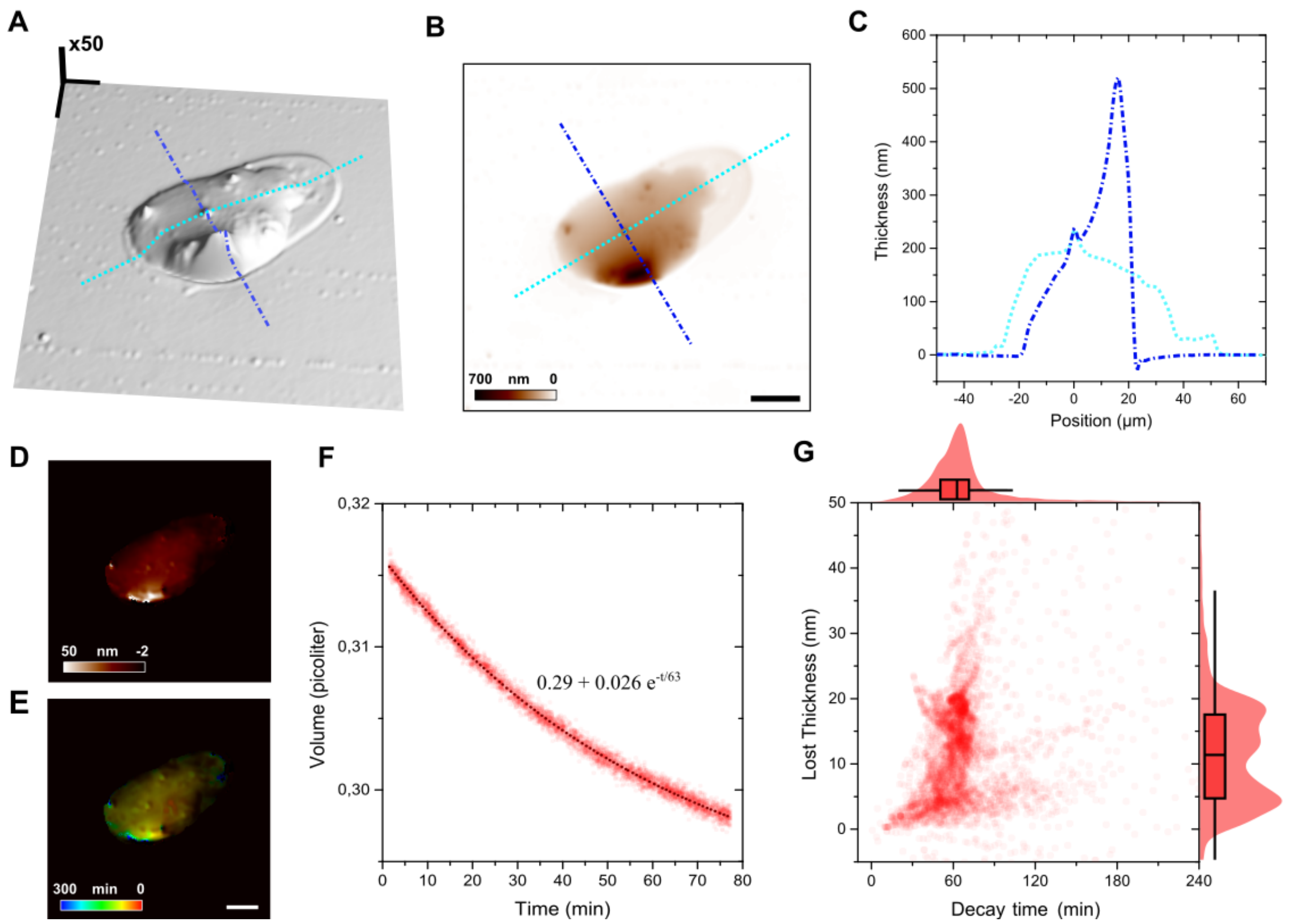

**Figure 2: Spatiotemporal dynamic of pheromone deposit at high resolution.** A trail pheromone deposit observed with quantitative phase imaging (a) 3d representation 1min after deposition, highlighting its heterogeneous thickness. The vertical axis has been amplified 50x for visualisation. (b) 2D map of a, the colour lookup table codes for the local thickness. (c) Thickness along the long (light blue dotted line) and short axis (dark blue dashed line) of the pheromone deposit (d) Map of the local fitted lost thickness in nanometres (exponential regression). (e) Map of the local fitted evaporation decay time in minute (exponential regression). (f) Long duration evaporation of the whole pheromone deposit with exponential fit.  (g) 2D plot of the lost thickness versus decay time in the pheromone deposit with the respective histograms and box plot (representing: <5%; 25%; 50%; 75%; 95%>). Scales bars (a-in plane, b): 20 µm. Scale bar (a-vertical): 400 nm

## Concordance between ant behaviour and direct imaging detection

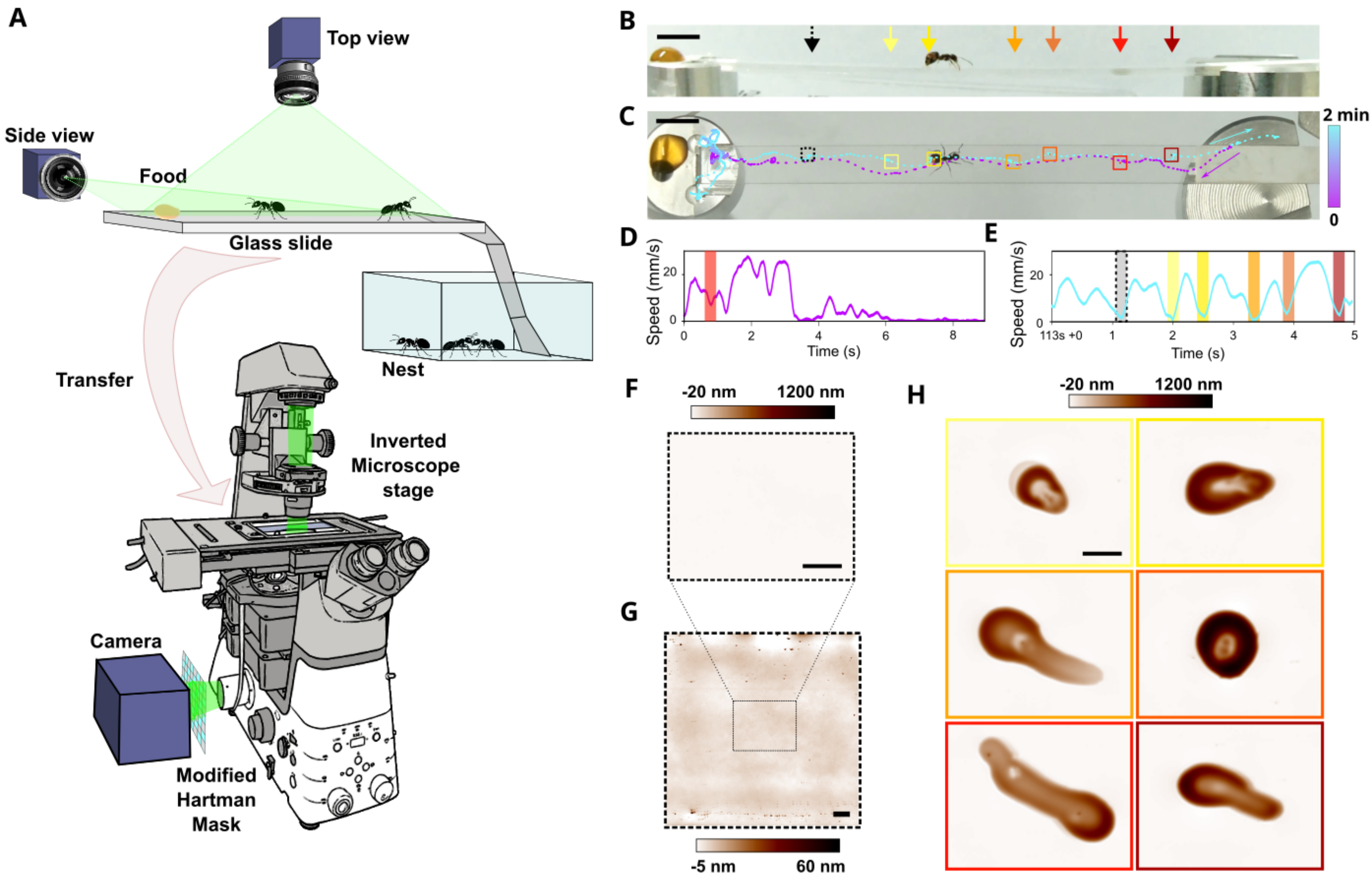

**Figure 3: Correlation between pheromone deposition behaviour and interferometric imaging**. (a) Schematic representation of the setup used in this part, 2 orthogonal cameras are used (top and side views) to observe the ant on the way to the food and back to the nest. A quantitative phase microscope equipped with motorized stage (on the back) is used to scan the whole glass slide. (b) Side view of an ant coming back from the food deposit. The arrows show the position when the ant displayed a pheromone laying behaviour (detected by eye from the movie recording of the ant). Solid arrows are used when we found a match between a laying behaviour and a deposit in the phase imagery scan. Dashed arrows are used when no pheromone deposit was observed. (c) Top view. The ant's abdomen trajectory is represented with the colour lookup table coding for the time. The boxes show the position when the laying behaviour was observed. Solid boxes are used when we found a correspondence between a laying behaviour and an actual pheromone deposit. Dashed boxes are used when no pheromone deposit was observed. The speed of the ant though time is presented, on the way to the food (d), and back to the nest (e). One behaviour was not correlated with a deposit in imagery (f) even when increasing the sensitivity by reducing the display dynamics (to detect thinner phase shifts) and increasing the size of the imaged region (g). The six other behaviours were associated with a detectable phase change (h, colour of the square contour matching the behaviours in panel b-e). Scale bar 5 mm (b,c) and 50 µm (f,g,h).

While literature had to rely on behavioural cues, these can be hard to detect and interpret, and are susceptible to observer bias. We compared behavioural observations with direct deposit imaging. *L. niger* pheromone deposition behaviour has been extensively studied, and has yielded significant insight on ant foraging mechanisms, such as the fact that they can modulate the number of deposits according to context, allowing the selection of the best food source (*18*) or the shortest path (*19*). *L. niger* deposit droplets intermittently, in a little Tom Thumb fashion. During deposition, the ant characteristically bends its abdomen towards the substrate, a behaviour commonly used to quantify pheromone laying and trail strength. We used two cameras to record the ant from the top and the side of the bridge (Fig 3a), and used the same protocol as in Fig 2. The top view was used for tracking the ant's position, and the side view to observe the deposition behaviour more easily. After a single ant completed her return trip to the nest, we imaged the entire bridge by stitching quantitative phase microscopic images (see methods). We used manual annotations of deposition behaviours observation in video recordings (arrows, Fig 3b). Deposits candidate areas are illustrated with coloured square boxes (Fig 3c). We also added an independent method to detect behaviours: speed drops from automatic tracking along the ant trajectory (Fig 3d-e), based on a study in the ant *Paratrechina longicornis* (*20*). While speed variations were frequently observed during a pheromone laying behaviour, similar speed drops also occurred for other reasons such as cleaning, resting or directional changes (Fig 3 d,e). In the representative ant shown in Fig 3, 7 putative deposition behaviours were observed (arrows, Fig 3b). Surprisingly, one of these behaviours (black dashed arrow and square, Fig 3 b,c) was not associated with any detectable phase change (Fig 3f), even looking over a larger field of view (700µm x 700µm) with increased sensitivity (reduced display dynamics, Fig 3g). This indicates that deposition behaviour does not systematically correspond to detectable deposits. We observed one deposit on the way to the food source (second arrow from the right Fig 3b,c) and 5 back to the nest (Fig 3h and see Suppl. Movie 3).

## Pheromone deposits variability

Our next step was to investigate the variability of single pheromone deposit. We again let a single ant walk on the glass bridge, then we immediately imaged the entire bridge on the microscope. The shape, homogeneity, surface, volume and circularity of the deposits vary both within and between ants (Fig 4 a-i, each ant is a different symbol). The deposit surface varied 10-fold with S=$4300^{+3400}_{-1200}$ µm² (Fig 4g). We then focused on the volume, as it represents the amount deposited. The variability was even higher (v=$0.68^{+2.3}_{-0.31}$ pL), with more than a 20-fold variation: the smallest deposit being less than a fifth of a picolitre, while the biggest was over 4 pL (Fig 4h). The deposits were round or pear-shaped, with varying elongation (circularity $0.49^{+0.11}_{-0.15}$, Fig 4i), likely due to the manner in which the ant dragged her abdomen on the substrate. A linear mixed-effects model with individual as a random effect revealed substantial inter-individual variability. The variance attributable to individual identity (3.36) was much greater than the residual variance (0.39), corresponding to an intraclass correlation coefficient (ICC) of approximately 0.90. This means that 90% of the total variance is explained by differences among individuals. It is worth noting that the observed differences are unlikely to result from remaining solvent, since the imaging is performed between 1 and 3 minutes after deposit (time to image the entire glass bridge on the microscope). These results challenge the common assumption that pheromone deposits are homogeneous in volume, at least in *L. niger*, and raise the possibility that some individuals play a disproportionate role in trail formation.

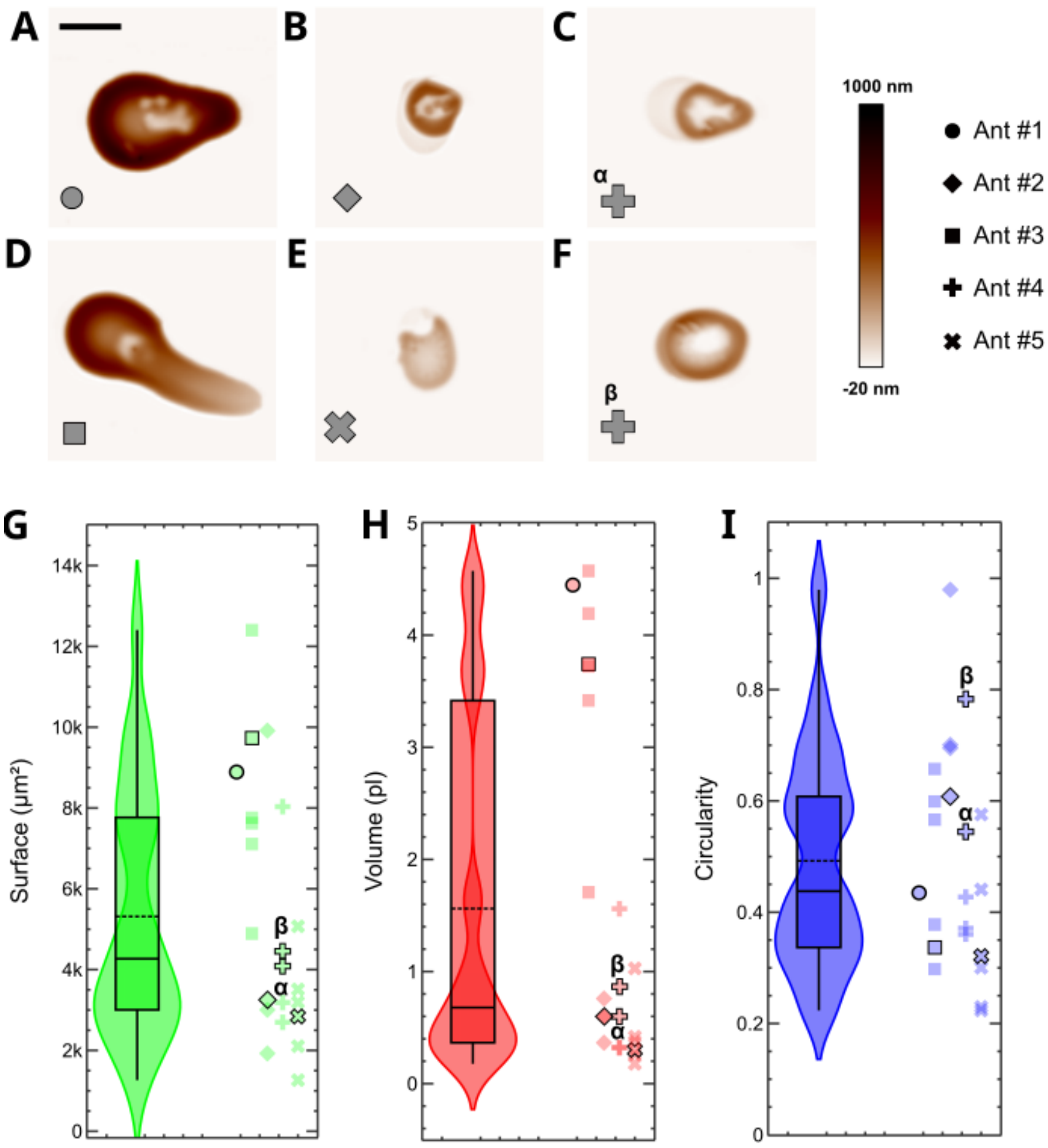

Figure 4: **Analysis of deposit morphology and variability**. (a-f) Images of pheromone deposit, the colour Lookup Table coding for the thickness. (g) Pheromone surface violin plot for 5 different ants, symbols on the right side represent single deposit, each symbol represent a different ant. (h) same as g for the total pheromone volume. (i) same as f for the pheromone circularity. Box plot (representing: <5%; 25%; 50%; 75%; 95%>, dashed line is the average), scale bar: 50 µm.

## Discussion

Our results bring the elementary building blocks of pheromone trail communication within experimental reach. Using quantitative phase imaging, a label-free, non-destructive method, we directly visualize and quantify individual ant pheromone deposits, revealing their morphology, volume and spatiotemporal dynamics over long time scales (> an hour) at video rate (25fps). Our approach directly detects pheromone deposits, bringing the elementary units of pheromone trail communication within experimental reach. Our approach avoids the false-positive and false-negative assignments associated with indirect methods, including the reference method based on behaviour detection. Although deposition behaviour mostly coincided with detectable deposits, this proxy was not fully reliable. Beyond detection, we found that individual deposits had volumes on the order of picolitres, with variable surfaces of 4300 µm$^2$ and a thickness of 134 nm (median values). Surprisingly, deposits varied greatly in shape, size and volume, both within and between individuals. On the glass substrate used in our experiments, a pheromone droplet followed two distinct behaviours. An initial rapid phase, lasting the first few seconds, corresponded to up to 80% of the total volume loss, and was followed by a much slower phase over approximately one hour, during which only 10% of the remaining volume was lost.

We thus observed three distinct physico-chemical behaviours within the deposits, suggesting the presence of three main components: a rapidly evaporating solvent constituting most of the initial volume; a sublimating solid characterized by slow mass loss and long persistence times; and a more stable solid residue persisting for several days. Gas chromatography coupled with massed spectrometry (GC-MS) methods—used to analyse pheromone composition by extracting components from an absorbing substrate after ants have deposited pheromones on it—therefore do not capture this first main component (solvent) of the pheromone deposits. We propose that the role of solvent is to reach a sufficient volume and fluidity for the ant to lay the pheromones. This organic solvent might be one of the major components of *L. niger* Dufour's gland (where pheromone are stored), liquid at room temperature such as undecane or decyl/dodecyl acetates (*21*). The sublimating solid is likely composed of isocoumarin. Indeed, this molecule is known to be an active molecule inducing trail following in *L. niger*. It was detected in GC-MS (*22*) and triggers a response in electro-antennography (*23*). Isocoumarin is an unstable solid at room temperature since its vapor pressure is 1 Pa (at 25°C) and is thus sublimating over time. The solid residue likely acts as a matrix to support the minute volume of the sublimating solid, and could promote chemical stabilization and resistance to environmental conditions. This solid state may also enable more precise spatial localization by limiting spreading and diffusion on the substrate.

Considering the pheromone deposit of Fig 2a, and making the hypothesis that the entire sublimated solid is isocoumarin, 1.1e11 molecules (26 picograms) would be released in the air. Considering the measured sublimation rate (63 min), an average of 240,000 molecules would thus be liberated per second from the deposit. Considering an average antenna effective surface (0.1 mm$^2$), 5cm away from the deposit, each antenna would receive an average of 100 molecules per second (see Suppl. Info. 2). Whether ants can detect these compounds from a single pheromone deposit remains unknown, as individual deposits could not previously be measured directly. Electroantennogram recordings using synthetic isocoumarin solutions showed detectable responses from 1 µg, a quantity perhaps exceeding that contained in a single natural deposit (*23*). However, these measurements were obtained from a dissected antenna, under artificial conditions using homogeneous liquid mixtures under an air flow, whereas natural trails consist of discrete deposits that may generate local gradients and allow active antennal scanning, potentially enabling detection at much lower quantities.

In the formicine ants the pheromone gland capacity tends to be less than 150 pg (*24*). If the entire volume of the deposit (solvent, isocoumarin and matrix) comes from the pheromone gland, this would mean that a single ant would have a limit of 100 deposits before the gland would need to be replenished. The rate of pheromone deposition, inferred from behaviour, typically ranges between 0.1 and 0.2 deposits per centimetre in *L. niger* (*13*). Combined with our volume estimates, this corresponds to a marking range of approximately 5 m, consistent with the known foraging range of the species. We found no data on pheromone replenishment dynamics, which, together with our approach, would allow quantification of the trade-offs between trail length, food quality and pheromone-laying capacity.

Our results provide experimental proof that the commonly used assumption of identical pheromone deposits is inaccurate, at least in *L. niger*. This is of importance since small differences at the individual level can have a disproportionate impact at the collective level (*25–27*). Evidence for repeatable inter-individual variation in insects is accumulating. In *Myrmica* ants, individuals showed repeatable levels of boldness and activity (*28*). Similarly, repeatability has been shown for exploratory behaviour in *Formica fusca* (*29*), foraging behaviour in *Tetramorium alpestre* (*30*), and tool use in *Aphaenogaster senilis* (*31*). If the differences we measured are repeatable, it would suggest the existence of specialist trail layers.

The large inter- and intra-individual variability observed in deposit further highlights the need to incorporate realistic pheromone dynamics into models of ant foraging and collective behaviour. In particular, evaporation rates are known to play a critical role in ant colony optimization (ACO) algorithms. For example, Mavrovouniotis et al. (*32*) showed that self-adaptive evaporation mechanisms outperform fixed or adaptive evaporation rates in dynamic optimization problems. By directly measuring the spatio-temporal dynamics of natural pheromone deposits, our approach provides experimentally grounded parameters for such models. An important next step will be to determine pheromone detection thresholds and the behavioural rules governing trail following. This remains challenging because detection itself is not straightforward. However, combining quantitative measurements of deposits with collective behavioural data should make it possible to infer detection and following thresholds experimentally. More generally, coupling our approach with techniques such as electroantennography and GC–MS could provide an integrated description linking chemical composition, physical dynamics and behavioural responses.

Beyond such fundamental questions, these measurements also have practical implications. Ants are among the most ecologically dominant animals, and several invasive species generate major economic costs worldwide. Their control often relies on poisoned baits exploiting collective foraging behaviour, and a better quantitative understanding of pheromone-mediated recruitment could improve the efficiency and specificity of these strategies. More broadly, pheromones are increasingly used as species-specific and environmentally friendly alternatives to conventional pesticides in agriculture (the global agricultural pheromones market is expected to reach $US 6.1 billion by 2026) (*33*).

In conclusion, the primary contribution of this study is the direct experimental access it provides to previously inaccessible communication signals. We establish the feasibility of label-free, real-time visualization and quantification of single ant pheromone deposits. This approach opens new possibilities for investigating how individual deposition events scale up to collective trail formation. It also brings currently unresolved aspects of ant decision-making within experimental reach, including potential context-dependent modulation of pheromone deposition and its associated trail-following behaviour. By providing direct localization and volume measurements, the method should also unlock chemical characterization of individual deposits. Finally, the approach can be extended to other insect species and environmental conditions, opening the way to systematic investigation of how phylogeny, ecology, and substrate properties shape the spatiotemporal dynamics of chemical communication.

## Acknowledgements

We thank the XLIM workshop for their help with the setup. We also thank Sahar Wehbi for fruitful discussions and support with the microscope.

*Funding statement:*
This work was supported by the CNRS and the University of Limoges. This project was supported by the French INSERM Booster Program "MECACELL3D". This project has also received funding from the European Research Council (ERC) under the European Union's Horizon 2020 research and innovation program (grant agreement No. [848645]), a grant from the University of Limoges, and a grant from the French National Research Agency (ANR-24-CE46-7295).

## Author contributions

L-A.P. and P.B. led the study. L-A.P. and P.B. initiated and conceived the research, with inputs from J.B. L.-A.P. conceived and conducted the ant behavioural experiments, while P.B. developed the imaging methodology. L-A.P., E.A and P.B. performed the experiments. L-A.P., T.L. and P.B. analysed the data and prepared the figures and videos. J.B. and P.B. performed the theoretical model for evaporation. T.L., L-A.P., and P.B. secured funding. L-A.P. wrote the original draft of the manuscript. P.B. substantially revised and edited the manuscript, and wrote the Supplementary section. All of the authors participated in discussions, data interpretation and manuscript editing.

## Competing interest declaration

The authors declare no competing interests.

## Additional information

Supplementary information is available for this paper.

Correspondence and requests for materials should be addressed to Laure-Anne Poissonnier laure-anne.poissonnier@unilim.fr and Pierre Bon pierre.bon@cnrs.fr.

Supplementary Materials for

# Imaging the building blocks of ant pheromone communication

**Authors**: Poissonnier L.-A.*, Amichot E., Brevier J., Liard T., Bon P.*
Affiliation: XLIM, CNRS UMR 7252, University of Limoges, Limoges, France
*: corresponding authors laure-anne.poissonnier@unilim.fr and pierre.bon@cnrs.fr.

**The PDF file includes:**

Material and Methods
Figures S1
Supplementary Text 1 to 2
Table S1

**Other Supplementary Materials for this manuscript include the following:**

Movies S1 to S3
Additional raw data on phase images are available on Zenodo (https://doi.org/10.5281/zenodo.20697871)

## Material & Methods

### Animal model

All experiments were carried with workers from a *Lasius niger* colony originating from a queen collected in a swarm in Toulouse in summer 2020. The colony had access to *ad libitum* water and sugar water (0.5M) and mealworms three times a week as a source of protein. The colony was kept at room temperature (20-30°C). The colony was starved 3-4 days prior to the experiments to ensure motivation to collect the food, as is classically done in such experiments.

### Experimental protocol

The experiment began by connecting the foraging area of the colony to a suspended glass bridge (57x4.5x1mm) by a metal ramp. To ensure cleanliness of the bridge and precise determination of the deposits, the bridges were cleaned in an ultrasonic isopropanol bath and inspected with high resolution quantitative phase imaging (see Suppl. Info. 3), and a different bridge was used for each ant. Ants were free to access the bridge, at the end of which they could feed on a 20 µL drop of 1M brown sugar water solution. To promote trail-laying and reduce stress effects, ants in parts 2 to 4 were allowed to explore the setup and eat before being tested. Ants that found the food and ate were marked with a dot of acrylic paint, and were the ones used for testing. Ants were always allowed to feed until satiation, and were placed in a temporary container (except in the live recording condition where they returned to their nest and free to come back) after they crossed the bridge to return to their nest (to prevent recording the same ant twice). The room temperature during the experiments was set between 25-26°C to maximise foraging activity.

### Live pheromone deposit quantitative phase imaging setup

To image the live pheromone deposit (Fig 1), we imaged a fixed, small portion (4mm x 4mm) of the glass bridge, which was illuminated from the top and at a wavelength around λ=525 nm with a collimated green LED (M530L4, Thorlabs) and a filter (FF03525/50, Semrock). An optical relay composed of two achromatic doublets ($f_1$=75mm followed by $f_2$=100 mm, Thorlabs) achieved a magnification of 1.3 between the glass bridge and a quadriwave lateral shearing interferometer (QLSI) (Bon *et al.* 2009) composed of i) a Modified Hartman Mask (MHM, Primot & Guerineau 2000) of period 24 µm, ii) an imaging relay (Moritex) and iii) a sCMOS camera (Dhyana 2100, Tucsen) to record self-interferences from the scene at up to 250fps and extract the quantitative phase image (see Suppl. Info. 3 for more details on the interference formation and numerical processing).

### High resolution quantitative phase imaging setup

To image pheromones at high resolution (Fig 2, 3, 4), a QLSI setup similar to the one presented in the previous method was placed at the exit of an inverted microscope (Ti2, Nikon) equipped with a motorized XY stage. This allowed to perform automized imaging of the whole glass slide using a stitching algorithm (ImageJ). The illumination was the native microscope LED filtered around λ=525 nm (FF03525/50, Semrock). In Fig 2, a 10x NA=0.25 achromat objective (Nikon) was used (resolution of 1 µm for a field of view of 340µm x 340µm recorded by QLSI) and in Fig 3,4, a 4x NA=0.1 achromat objective (Nikon) was used (resolution of 2.6 µm, field-of-view of 850µm x 850 µm recorded by QLSI).

Dual view imaging for tracking and behaviour observation

In Fig 3, the glass bridge was recorded simultaneously from the top and the side with two Raspberry Pi cameras equipped with video objectives (3MP 8-50mm for top view and 3MP 6mm for side view). The movies were recorded at 25 fps on Raspberry Pi 4b micro-computers.

Tracking

Ant positions in each frame of the top-view recordings were extracted using the software AnimalTA (Chiara & Kim 2023). In the rare cases where automatic detection failed, positions were manually annotated by clicking on the ant location. All trajectories were visually validated by overlaying coordinates onto the original recordings. Speeds were calculated from the distance travelled between successive frames and averaging over 5 frames.

Phase measurement interpretation

Quantitative phase imaging using QLSI records the wavefront modification impinged by the sample, also called Optical Path Difference (OPD). Under imaging condition between the sample and the sensor and pseudo-collimated trans-illumination, the OPD can be described as $OPD = \Delta n \cdot e$ , with $\Delta n$ the local refractive index difference between the sample and the surrounding medium (air in our case, *n* = 1) and *e* the local thickness. In this publication we assume $\Delta n = 0.5$ as constant, neglecting the variation of refractive index between the different chemical components, to retrieve the thickness from the OPD. This assumption is justified since most of the organic compounds have a refractive index scaling from 1.4 to 1.6 (e.g. undecane $n_{und}$ = 1.42, dodecyl acetates $n_{\mathrm{dod}}$ = 1.43, sucrose $n_{sucr}$ = 1.57, coumarin $n_{\mathrm{cou}}$ = 1.51). This means that the calculated thicknesses uncertainty is about 20%.

Statistical analysis

To quantify variation in deposit volume, while accounting for repeated measurements from the same individual, we fitted a linear mixed-effects model with deposit volume as the response variable and ant identity as a random effect using library lme4 (Bates *et al.* 2015) in RStudio version 2026.04.0+526 (R core team 2024).

Methods references

**Figure S1: Observation of pheromone deposit residue 24h hours after deposit.**

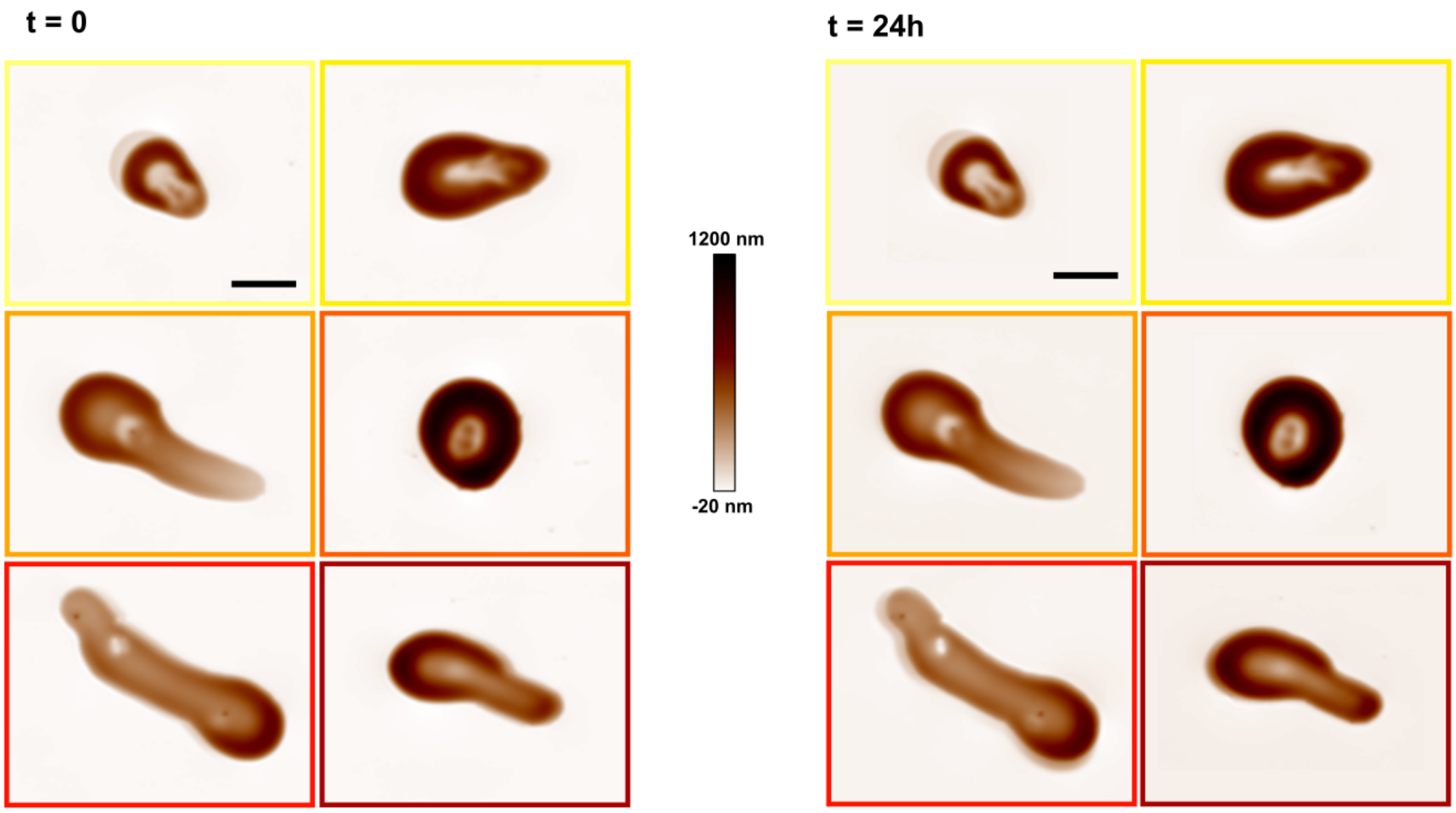

*Fig S1. Pheromone deposit 1 minute after deposit (left) and 24h after deposit (right). The colour lookup table code for the thickness. Scale bars: 50µm*

**Supplementary Texts**

1) Sublimation and diffusion of pheromones

Isocoumarin is known to be present in pheromone trails of *Lasius niger*, detected and followed for around an hour after initial deposit [1,2]. Isocoumarin melting point (47°C) is significantly above room temperature and it thus tends to sublimate since its vapor pressure is $P_{sat} \approx 1$ Pa (at 25°C). After the fast evaporation of the solvent (within few seconds: Fig 1), let us consider that the matter loss in pheromone deposit (measured with quantitative phase imaging) is only due to isocoumarin sublimation. Isocoumarin density is evaluated from isocoumarin isomers[3] to be $d\sim 1$ g·cm$^{-3}$. Its molar mass[3] is $M=146.14$ g·mol$^{-1}$ and its refractive index is n = 1.59 (close to the average value of 1.5 taken to convert optical path difference into thickness). From the data of Fig 2, we measure an evaporation of $v=26$ fL with a time constant of $\tau_{long}=63$ min. We thus obtain a total number $N$ of evaporated isocoumarin molecules of:

$$N=\frac{v\cdot d\cdot Na}{M\cdot}=1.1\cdot 10^{11} \text{ molecules}, \qquad \text{(eS2.1)}$$

with Na the Avogadro number. Considering, as first approximation, that the sublimation lasts $2\cdot\tau_{long}$ with a constant rate (due to constant surface of the solid deposit), we then determine the sublimation rate $S_N$ to be:

$$S_N=\frac{N}{2\cdot\tau_{long}}=1.5\cdot 10^{7} \text{ molecules}\cdot \text{s}^{-1}. \qquad \text{(eS2.2)}$$

As presented in Fig S2, we consider the molecules diffusing in a half space as the pheromone is deposed on an infinitely small surface. If we suppose the steady state of molecule diffusion, a constant flow of molecule $\vec{\Pi}$ will travel through each elementary surface dS.

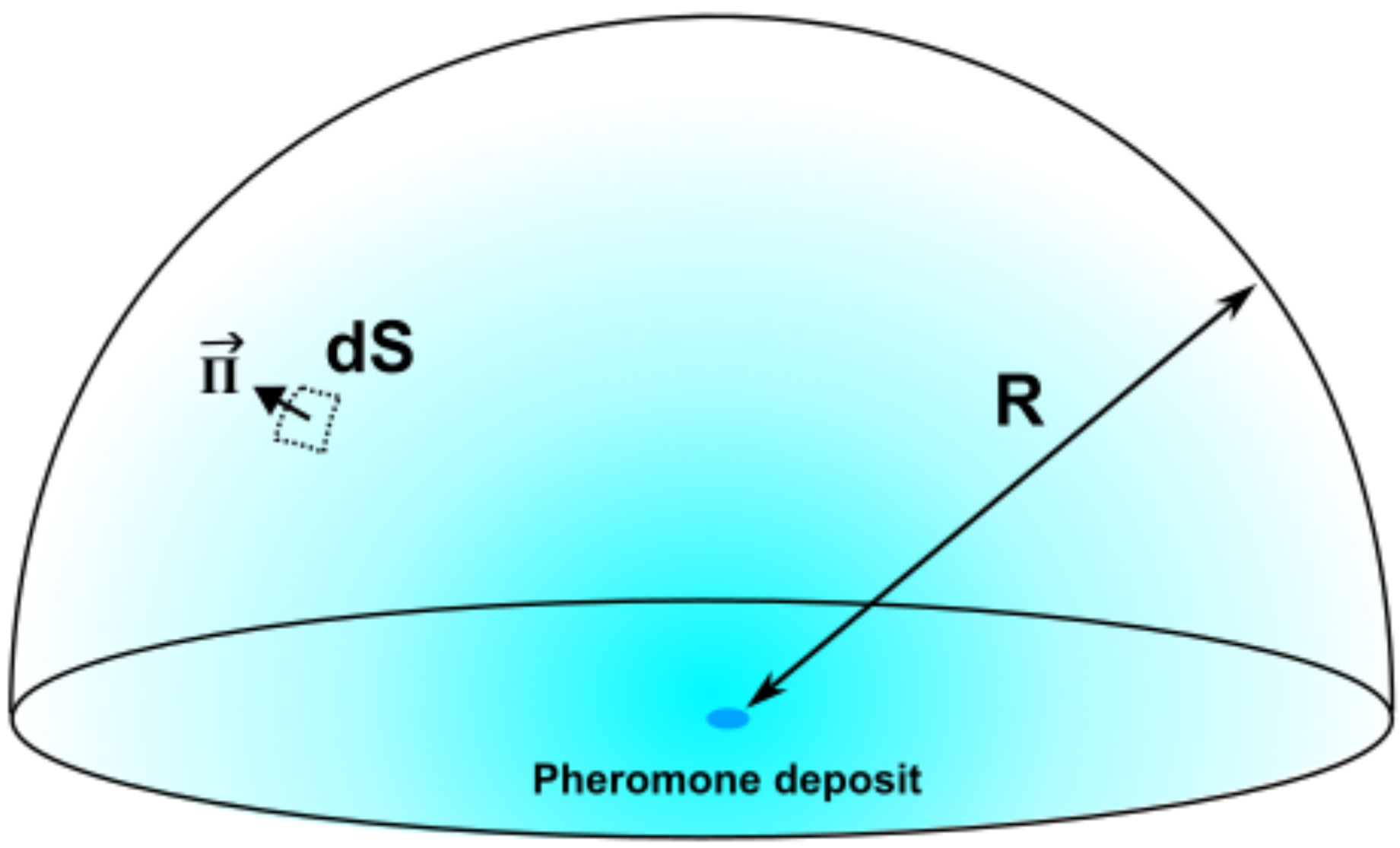

*Fig S2. Scheme of the diffusion of pheromone deposit*

At a distance $R$ from the deposit, we obtain thus a flow of:

$$\|\vec{\Pi}\|=\frac{S_N}{2\pi R^2} \text{ molecules}\cdot \text{m}^{-2}\cdot \text{s}^{-1}. \qquad \text{(eS2.3)}$$

Considering an ant detecting the pheromone deposit at a distance of 5 cm, the flow would be $\approx 10^9$ molecules·m$^{-2}$·s$^{-1}$. Considering the antenna projected surface as rectangles ($S_{antenna}$=1 mm x 0.1 mm = 0.1 mm²) intercepting the flow, we obtain around 100 molecules·s$^{-1}$ reaching the surface of each antenna.

2) Quantitative phase imaging using quadriwave lateral shearing interferometry.

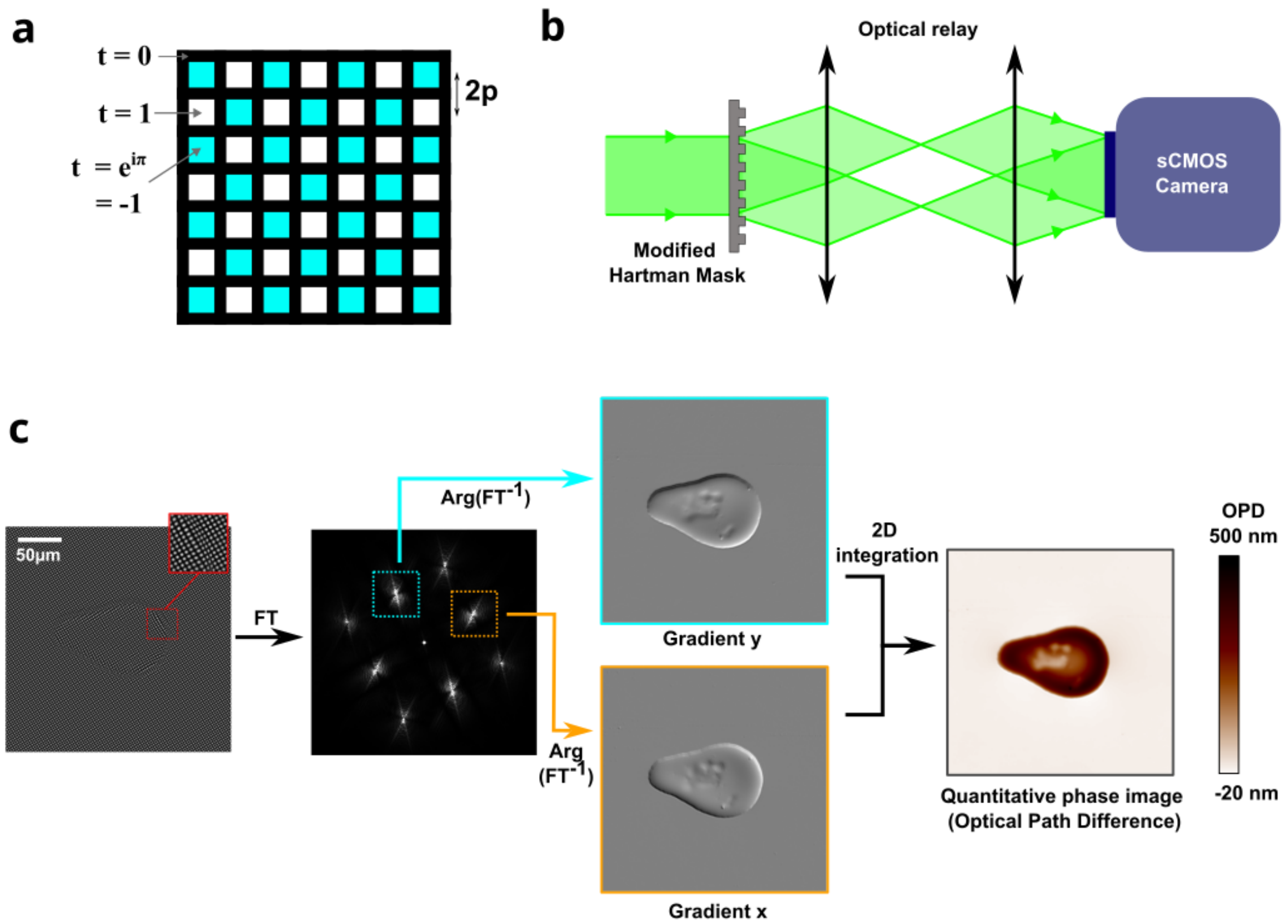

*Fig S3. (a) Modified Hartman Mask and transmittance. (b) scheme of a quadriwave lateral shearing interferometer. (c) Processing pipeline for quantitative phase measurement*

The quantitative phase measurements are based on the analysis of the self-interferences generated by a two-dimensional diffraction grating. The grating used (modified Hartman mask, MHM[1], Fig S3.a) is designed to generate 4 replicas of the incoming wavefront : 2 along the x direction, 2 along the y direction. The perfect transmittance to perform this generation can be written as:

$$t_{perfect}\,(x, y) = \cos\frac{\pi x}{p} \cdot \cos\frac{\pi y}{p}, \qquad \text{(eS3.1)}$$

with $p$ the half period of the grating. In practical, its transmittance $t$ is a 3-level approximation of equ.S3.1 and is obtained by combining an amplitude grating of period $p$, usually *via* deposit of metal producing region with $t$=0, and a phase grating of period 2$p$, usually by etching of glass producing a chessboard pattern with an alternance of $t$=1 and $t$=-1

After a small propagation of $z$, the replicas interfere on a 2D sensor and generate an interferogram $I$:

$$I(x,y) = i_0\left\{1 + \cos\left(\frac{2\pi}{p}\left[x + z\frac{\partial \mathrm{OPD}}{\partial x}\right]\right) + \cos\left(\frac{2\pi}{p}\left[y + z\frac{\partial \mathrm{OPD}}{\partial y}\right]\right) + \frac{1}{2}\cos\left(\frac{2\pi}{p}\left[x + y + z\frac{\partial \mathrm{OPD}}{\partial (x+y)}\right]\right) + \frac{1}{2}\cos\left(\frac{2\pi}{p}\left[x - y + z\frac{\partial \mathrm{OPD}}{\partial (x-y)}\right]\right)\right\},$$

(eS3.2)

with $i_0$ the beam intensity. In this publication, we have used on optical relay between the MHM and the sensor to simplify the mechanical alignment.

The optical path difference (OPD) is encoded in the interferogram *via* frequency modulation at frequency $1/p$ and $\sqrt{2}/p$. The algorithm used in this paper perform a numerical Fourier transform of the raw interferogram to allow a crop around the two peaks at $1/p$. A numerical inverse Fourier transform is applied to each Fourier subimage to obtain demodulated harmonics:

$$\begin{cases} H_x = i_0 e^{i\frac{2\pi}{p}z\frac{\partial \mathrm{OPD}}{\partial x}} \\ H_y = i_0 e^{i\frac{2\pi}{p}z\frac{\partial \mathrm{OPD}}{\partial y}} \end{cases} \quad \text{(eS3.3)}$$

After getting the angle of each harmonic, the two orthogonal OPD gradients $\frac{\partial \mathrm{OPD}}{\partial x}$ and $\frac{\partial \mathrm{OPD}}{\partial y}$ can be integrated in 2D using complex valued integration [2] to obtained the OPD.

To remove the contribution from everything but the sample itself, a reference image is taken on a region which can be considered as empty and subtracted to the OPD gradients before integration.

Other quantitative phase methods have been developed with different advantages, limitations and phase retrieval algorithms[3].

**Table S1.**

Single deposit data. In the first column is the deposit number. In the second the glass slide number (corresponding to a single ant, an ant could make several deposits on the same slide). The third column is the volume of the deposit in picoliters. The forth column is the surface in micrometers$^2$. The last column is the ratio between the fitted elliptical short and long axes.

| Deposit ID | Slide ID | Volume (pL) | Surface (um2) | Small axis / Large axis |
|---|---|---|---|---|
| 1 | 3 | 4,45 | 8877,22 | 0,56 |
| 2 | 7 | 3,72 | 9730,73 | 0,34 |
| 3 | 7 | 3,42 | 7622,93 | 0,38 |
| 4 | 7 | 3,74 | 7765,74 | 0,57 |
| 5 | 7 | 1,71 | 4892,50 | 0,60 |
| 6 | 7 | 4,57 | 7103,12 | 0,66 |
| 7 | 7 | 4,19 | 12404,00 | 0,30 |
| 8 | 8 | 0,60 | 9913,52 | 0,98 |
| 9 | 8 | 0,76 | 3004,62 | 0,70 |
| 10 | 8 | 0,60 | 3244,53 | 0,61 |
| 11 | 8 | 0,37 | 1925,01 | 0,69 |
| 12 | 17 | 1,56 | 8034,21 | 0,37 |
| 13 | 17 | 0,33 | 2696,16 | 0,43 |
| 14 | 17 | 0,60 | 4087,08 | 0,54 |
| 15 | 17 | 0,32 | 3184,55 | 0,36 |
| 16 | 17 | 0,87 | 4452,66 | 0,78 |
| 17 | 0 | 0,30 | 3191,06 | 0,22 |
| 18 | 0 | 0,26 | 2105,29 | 0,58 |
| 19 | 0 | 0,38 | 2840,11 | 0,30 |
| 20 | 0 | 0,18 | 1263,54 | 0,23 |
| 21 | 0 | 1,03 | 5070,61 | 0,32 |
| 22 | 0 | 0,42 | 3510,49 | 0,44 |

**Movie S1.**

Quantitative phase imaging of ants and trail pheromone deposits during 3min40s at 20 fps. Two pheromone deposits were observed from different ants.

**Movie S2.**

Time lapse of pheromone deposit evolution. Left: Total thickness from quantitative phase imaging. Right: Thickness variation from the beginning of the movie.

**Movie S3.**

(up) Top and side view of an ant walking back from a food source. Different pheromone deposit behaviours are visible. (down) Quantitative phase imaging converted to thickness in the regions where the behaviour was observed. The first behaviour was not linked to any deposit.